\documentclass{article}
\usepackage{latexsym,amsfonts,amssymb,  
}
\begin{document}
%\begin{verbatim}
 
\title{Computability at the Planck Scale}

\author
{
Paola Zizzi \\
Dipartimento di Matematica Pura ed Applicata,
\\
Universit\`a di Padova
\\
via G. Belzoni n.7, I--35131 Padova, Italy}

\date{}
\maketitle

\begin{abstract}
We consider the issue of computability at the most fundamental level of physical reality: the Planck scale. To this 
aim, we consider the theoretical model of a quantum computer on a non commutative space background, which is a 
computational model for quantum gravity.

In this domain, all computable functions are the laws of physics in their most primordial form, and non computable 
mathematics finds no room in the physical world.

Moreover, we show that a theorem that classically was considered true but non computable, at the Planck scale 
becomes computable but non decidable. This fact is due to the change of logic for observers in a quantum-computing 
universe: from standard quantum logic and classical logic, to paraconsistent logic.
\end{abstract}

A quantum computer \cite{NC} can simulate \cite{F} perfectly and efficiently a quantum mechanical system.
Precision is due to discreteness from both sides: qubits in the quantum
register, and discrete spectra in the quantum system. 
Efficiency is due to quantum parallelism in the quantum computer.
However, there is not an isomorphism between the quantum computer and the quantum system, because the latter lies on 
a classical background (classical space-time, which is a smooth manifold).

The classical background is not taken into account during the simulation, but only at the end, when a classical 
output is obtained by measurement. Because of that, a great deal of quantum information remains hidden. 
The hidden quantum information is due to the irreversibility of a standard quantum measurement, but this is related 
to the situation of the external observer in a classical background, which cannot be put in one-to-one 
correspondence with the machine computational state. 

One might argue, then, that in a discrete space-time like a lattice, one could overcome this problem. However, we 
wish to recover the hidden information in a physical discrete space-time, that is, both discrete and Lorentz 
invariant. And, obviously, a lattice breaks Lorentz invariance. 
The simplest way out is to consider a quantum space background like a fuzzy sphere \cite{Ma}, which is both discrete and 
Lorentz invariant. This choice arises quite naturally. In fact, in a recent paper \cite{Z}, we showed that the non 
commutative algebra of quantum logic gates of a quantum computer of N qubits, is associated, by the non-commutative 
version of the Gelfand-Naimark theorem \cite{KW}, with a quantum space which is a fuzzy sphere with $n=2^N$ elementary cells.

The background space and the quantum computer are then in a one-to-one correspondence. An "external" observer on 
the quantum space background automatically becomes an "insider" observer with respect to the quantum computer.
Thus, in this model, one can conceive a reversible quantum measurement \cite{Z}, performed by an "insider observer", a 
hypothetical being living on the fuzzy sphere. There is no hidden information anymore: all the quantum information 
remains available. 

Recently, we investigated about the logic \cite{BZ} of the internal observer, which turns to be a paraconsistent 
\cite{DC}, \cite{P} 
and symmetric logic, like basic logic \cite{SBF}, where both the non-contradiction and the excluded middle principles 
are invalidated.
The isomorphism between the quantum computer and the quantum space background turns to be quite useful, as it 
leads to a quantum-computational basic formulation \cite{Z2} of quantum gravity (space-time at the Plank scale).

It is usually assumed that, at the Planck scale the very concept of space-time becomes meaningless, and should be 
replaced by a discrete geometrical structure. (A major example is Loop Quantum Gravity (LQG) \cite{R}, \cite{T}, 
\cite{Pe}, where the 
discrete structure is realized by spin networks \cite{Pen}, \cite{RS}).

We believe that the modification of the concept of space-time at the Planck scale leads to a modification of the 
concept of computability.
(This does not mean, however, that at the Planck scale there is the so-called hyper computability 
\cite{Ho}, \cite{Ki}, \cite{Cap}, \cite{Cal}, by which 
it would be possible to construct some hyper-machines, or utilize some quantum systems that can calculate 
non-recursive functions). 

It might be that the quantum hidden information, due to the inconsistency between Quantum Mechanics and its 
classical background, is responsible of the appearance of non computable problems, in relation with the standard 
quantum logic \cite{BvN} of the external observer.
In fact the proof of a theorem can be considered as an algorithm, and when this algorithm is run by a quantum 
computer, you can only know if it is true or false but the proof will remain unknown to you. 
Then, one is faced with the same kind of problems encountered in the simulation of a quantum system.

Instead, if one imagines a quantum computer embedded in a quantum space or, in other words, a "quantum computing space" 
(the classical world having been "erased"), the very definition of computability is modified. This is related to the new 
logic of the "internal" observer.
In fact, if an "observer" could enter a quantum space and put itself in a one-to-one correspondence with the quantum states 
of the computer, then a complete meta-system would arise. 

In other words, in a quantum space-time isomorphic to a quantum 
computer, all problems would appear to be computable. However, the complete meta-system would appear "inconsistent" 
with respect to a classical world. Of course the Planck scale is (at least at present) unreachable experimentally, 
nevertheless, studying quantum computability at the fundamental scale, can lead to a deeper insight into the foundations 
of Quantum Mechanics.

Deutsch \cite{De}, \cite{De2} claims that the laws of Physics determine which functions can be computed by a universal computer. Also, he says that the only thing that privileges the set of computable functions (or the set of quantum-computable functions) is that it is instantiated in the laws of physics.
But one can go a bit further: in the quantum computer view of space-time at the Planck scale \cite{Z3}, quantum space-time is a universal quantum computer that quantum-evaluates recursive functions which are the laws of Physics in their most primordial and symbolic form. In other words, at the Planck scale because of the isomorphism between a quantum computer and quantum space-time (quantum gravity), the laws of physics are identified with quantum functions. This is the physical source of computability, and leads to the conclusion that at the Planck scale, only computable mathematics exists.

We would like to make a remark: Deutsch says that all computer programs may be regarded as symbolic representations of 
some of the laws of physics, but it is not possible to interpret the whole universe as a simulation on a giant quantum 
computer because of computational universality. We fully agree with that, and we wish to make it clear that, in our view, 
quantum space-time is not a simulation but is itself a quantum computer, and, by quantum evaluating the laws of Physics, 
it just computes its own evolution.

One might then ask what would happen to the observer logical dichotomy mentioned above, in the case the whole universe 
were a giant quantum computing-fuzzy sphere.
Actually, by using the holographic principle \cite{Hoo}, and the isomorphism between the quantum computer and the fuzzy sphere, 
one is able to settle a minimal model \cite{Z2} for Loop Quantum Gravity, which has been called Computational Loop Quantum Gravity (CLQG).
This leads to a discrete area spectrum for the cosmological horizon, which can be viewed as the surface of a giant fuzzy sphere, whose elementary cells encode the whole quantum information of the universe.

The amount of quantum information stored by the cosmological horizon since the Big Bang was computed in \cite{Z4}, 
\cite{Z3} 
and \cite{Ll}, and it turned to be  $N\approx 10^{120}$ qubits.
The area of an elementary cell [9] of the event horizon of a fuzzy black hole-quantum computer is: 
$$
A_{EC}=\frac{N}{2^N}A_0,
$$
where $A_0$  is the elementary area in LQG, of the order of a Planck area: $A_0\approx l^2_P$ , where  
$l_p\approx 10^
{-33} cm$ is the Planck length. In the case the whole universe were a fuzzy quantum computer, an elementary cell of 
the cosmological horizon would have an area: 
$$
A_{EC}=\frac{10^{120}}{2^{10^{120}}}A_0,
$$
which is extremely small, 
but finite.

If the internal observer, inside the fuzzy quantum universe, is still able to intuitively recognize discreteness of 
space-time, he will also be able to use paraconsistent logic in his judgments. If instead the observer focuses on 
his perceptions, he will make, in his mind, automatically the two limits: $N\longrightarrow \infty$ and $A_{EC}
\longrightarrow 0$. 
In this way he would pretend he lives in a classical space-time, and will automatically become an Aristotelian 
logician, with respect to the universe as a whole. It should be noticed that those limits are full of implications. 
In fact, they mean that an infinite string of bits (infinite classical information) is encoded in a point of space. 
We would like to call this situation: "Information singularity" as it reminds the Big Bang, where an infinite 
amount of energy (instead of information) is concentrated in a point.

However, there is effectively an increase of information entropy, in this model, as far as $N$ increases, as well 
as a tendency to the continuum.
We are faced with signals of higher complexity and quantum information loss. It becomes harder and harder, for the internal observer, in this situation, to maintain a coherent superposition of truth values in his logical judgements. In fact, he starts to lose his capacity to distinguish the discrete space-time structure from a smooth manifold, and perceives an increase of complexity. 
We argue then, that non computable mathematics is an emergent feature of computable mathematics at higher levels of complexity, as perceived by the 
observer/logician.
In digital physics \cite{Fr}, \cite{Wo}, which relies on the strong Church-Turing thesis (the universe is equivalent to a Turing machine) it is assumed that it exists a program for a universal computer which computes the dynamical evolution of the universe. 

Of course, this is an algorithmic view of physical reality, and it can be easily adopted if one assumes that real numbers and continuous physical quantities comprised continuous space-time, are absent. Because of that, when the limits discussed above are taken into account, the digital paradigm starts to fade, and needs some extra insights. We believe that the relation between non computability and emergent information 
complexity might be better investigated in the context of Chaitin's algorithmic information theory (AIT) \cite{Ch}. 

In our model, as we have seen, the universe is the result of the identification of a huge quantum computer with a quantum (non commutative) space.
The software version of the cosmic computer represents the universe as a simulation. But we, like Deutsch, strongly disagree with this approach. Instead, in the hardware version, the universe computes its own dynamical evolution. The cosmological models described in 
\cite{Z3}, \cite{Z4} and particularly in \cite{Z2}, where LQG is itself the cosmic computer, obviously belong to the hardware version. 
As Deutsch illustrated in \cite{De3}, a universal quantum Turing machine (or quantum computer) has the same computability power of a universal classical Turing machine.

It follows that, as a quantum computer, the universe can compute only Turing-computable functions. As a physical universe, it computes the laws of physics in their most primordial form.
Then, the laws of physics are all Turing-computable. Consequently, non computable mathematics is outside the realm of the physical world. 

The observer is internal with respect to the fuzzy quantum-computer universe, and can only use 
paraconsistent logic, with respect to the universe as a whole. In the mind of the observer, there are two strong axioms, which are the converse of the non contradiction and excluded middle principles, respectively.
The first axiom, namely the converse of the non contradiction principle: $\vdash A\land \neg A$, does not 
allow the observer to use classical truth values in his judgements (measurements) and, for him, classical decidability 
is now meaningless. 

However, the internal observer is also external with respect to a quantum subsystem, like a quantum computer inside the quantum computing universe. In this case, the observer utilizes standard quantum logic. The Boolean outputs of his standard (irreversible) quantum measurement are the truth values of classical logic.
The double logic of the observer, leads to a change in the interpretation of the first G\"odel's incompleteness theorem \cite{Go} 
at the Planck scale, as we will see in the following. Simply, there is a change in the concept of truth: what classically was true but non-computable appears now to be computable but non decidable.

Let us consider the famous "couple" of quantum computing: Alice (A) and Bob (B). 
Alice (A) lives in a classical space-time, which is also the classical background of the quantum computer (QC). A is an external observer with respect to the QC, and is equipped with standard quantum logic when she performs a standard quantum measurement; otherwise, she is endowed with classical logic.
Instead, B lives in a quantum space isomorphic to the QC, he is an internal observer, and is equipped with paraconsistent logic.

Let us suppose that a theorem (T) is given to A. 
T is computable, but the computation of T is too long (and hard) for A and for her classical computational tools (classical computer or classical Turing machine).
Then, A gives T as an algorithm to the QC. By using quantum parallelism (and entanglement), the QC computes T efficiently.
T is true, and the output is $1$. However, A will never have the demonstration at her disposal, because, to get the result, she had to destroy the superposition, and lose quantum information.
Thus, T appears to A as G\"odel's formula G would have appeared to a mathematician: true, but not (for A) computable.
Of course, T "is" computable, as the QC has just computed it. But, with respect to the external observer (A), the QC has behaved like another human (B), the internal observer. In fact B can verify that T is computable, but will tell A that T is not computable, as he will never provide her the demonstration.

Then, A will enter in a quantum space whose elementary cells are in a one-to-one correspondence with the computational 
states (B's states).
Now, A can verify herself that T is computable, as B is computing it.
But now, neither A nor B will ever know whether T is true or false, because $0$ and $1$ are classical truth values, which are not obtainable anymore, now that the classical world has been "erased" for the couple A-B.   

Let us suppose that the QC is the whole universe. Now A and B are both internal observers, and both use paraconsistent 
logic with respect to the universe as a whole.  The question: "Is T is true or false?" has no answer. 
In the passage from the classical to the quantum stage, there is a change in the meaning of definitions.

In the classical setting, T was true, but the quantum demonstration was not provided, so T appeared as G\"odel's sentence 
G, true but non computable.  In the quantum setting, the quantum demonstration of T is available: T clearly appears to be 
computable. But the classical truth values are now meaningless: T is (classically) undecidable.

\medbreak
{\bf Acknowledgements}
I wish to thank G. Battilotti and G. Sambin for useful discussions. 
Work supported by the research project "Logical Tools for Quantum Information Theory", 
Department of Pure and Applied Mathematics, University of Padova.

\end{document}